\documentclass[aps,prb,twocolumn,epsfig,floatfix,twoside,a4paper,showpacs]{revtex4}
\usepackage{epsfig}

\begin{document}
\input epsf
\title{Structural properties and liquid spinodal  
of water confined in a hydrophobic environment.}
\author{
P.~Gallo$^\dagger$\footnote[1]{Author to whom correspondence  
should be addressed; e-mail: gallop@fis.uniroma3.it} and
M.~Rovere$^\dagger$ }
\affiliation{$\dagger$ Dipartimento di Fisica, 
Universit\`a ``Roma Tre'', \\ 
and Democritos National Simulation Center,\\ 
Via della Vasca Navale 84, 00146 Roma, Italy }

\begin{abstract}
\noindent
We present the results of a computer simulation study of 
thermodynamical properties of TIP4P water confined in a hydrophobic 
disordered matrix of soft spheres upon supercooling. 
The hydrogen bond network of water appears
preserved in this hydrophobic confinement.
Nonetheless a reduction in the 
average number of hydrogen bonds due to the geometrical constraints
is observed.
The liquid branch of the spinodal line is calculated from 350 K
down to 210 K. The same thermodynamic scenario of the bulk is found:
the spinodal curve is monotonically decreasing. 
The line of maximum density bends avoiding a crossing of the spinodal.
There is however a shift both of the line of maximum density and
of the spinodal toward higher pressures and 
lower temperatures with respect to bulk.

\end{abstract}

\maketitle

\section{Introduction}

The behaviour of the liquid-gas spinodal in water
plays an important
role in the interpretation and the
determination of its complete phase diagram. In particular
the related behaviours of the liquid branch of the spinodal and the line of 
maximum density (TMD line) in the supercooled liquid state are 
connected to the possible existence 
of a second critical point~\cite{gene1,gene2}.
Among the different interpretations of the anomalies of
water at low temperatures in fact a number of experimental and
computer simulation results and theoretical calculations
indicate the possibility of a liquid-liquid (LL) coexistence
and a second critical point~\cite{pabloreview,genepablo}. The
low density and high density liquid phases would be
in correspondence with the 
low density and high density amorphous forms of water below
its glass transition temperature. 
In accordance with these hypotheses 
the many numerical studies performed to determine the spinodal line
of bulk supercooled water agree in finding that this line
is monotonically decreasing at least down to
the lowest temperatures that was possible to 
investigate~\cite{spinodal1,Essmann,poole,tanaka,hpss,netz,Mossa,minozzi}.
Correspondingly the TMD line bends to avoid crossing the spinodal.
This scenario excludes both TMD and spinodal behaviour to be
responsible of the water singularities. These singularities
can therefore be connected to the existence of the second
critical point or to a singularity free scenario~\cite{sastry-singfree}.

An increasing amount of theoretical and experimental studies have been 
performed in recent years on water in confined 
geometries, see for example~\cite{Rossky,Marti,Starr2,Rossky2,Swenson,Crupi,SowHsin,starr,Pellenq,SowHsin2,SowHsin3,gallo1,ricci,torquato,rovere,gallo2,meyer,koga}. 
Water is in fact
confined in many situations of interest
for biological and technological applications. In this respect
the study of the
modification of the phase diagram of water induced by the perturbation of
the confining environments is particularly significant.
The study of water in confinement also can help to shed light on the
behaviour of the bulk in the region of deep supercooling
were nucleation prevents experiments to be performed in the
bulk while permits very low temperatures to be reached in 
confinement\cite{SowHsin,SowHsin2,SowHsin3}
displaying scenarios that can be connected to the bulk 
behaviour\cite{PNAStutti}.
The shift of the phase diagram due to confinement could drive the so far
experimentally inaccessible zone where the second critical point is
supposedly located in a region accessible to experiment.
A study of ST2 water confined between smooth plates has indicated
the possibility of a LL transition in confinement~\cite{meyer}.
A recent theoretical work indicates a shift to lower temperatures,
higher densities and higher pressures of the second critical point
for hydrophobic confinement~\cite{torquato}.
Indications of a LL transition have been also found in a recent
computer simulation study~\cite{starr} on TIP5P water confined between 
hydrophobic plates,
a shift to lower temperature of the second critical point is also estimated.

Generally speaking solid interfaces strongly distort the HB
network~\cite{torquato}. In some cases 
an hydrophobic environment for water can be provided by the presence of
an apolar solute, as in the case of aqueous solutions of rare gases,
polymers or large organic molecules. 
When the hydrophobic units are small enough 
it is expected that water
maintains its hydrogen bond structure~\cite{chandler,chandlernat} 
in spite of the
perturbation of the solute, this leads to 
the formation of water cages around the non polar solute. In this
paper we explore how this hydrophobic effect influences
the behaviour of the spinodal line and the TMD upon cooling
in the limit of small hydrophobic units to study the case where the
structure of water is preserved or only slightly modified.
Since the solute-solute interaction is expected not to play an
important role in comparison with the hydrogen bond 
and the solute-solvent interactions, in the present study we consider
as a first approximation a model where few solute particles are kept fixed to
form an hydrophobic confining matrix for water. 
We expect this model to catch the main features of the
phenomenon.

\section{Model and computer simulation details}

The water fluid is simulated with the DL\_POLY package 
(W. Smith, T. R. Forester
and I. T. Todorov, Daresbury Laboratory, UK) 
using the TIP4P site model~\cite{tip4p}. 
In this model 
three sites are arranged according to the molecular geometry. The
two sites representing the hydrogens are 
positively charged with $q_H=0.52$ atomic charges, each one forms a rigid bond
with the site of the oxygen at distance
$0.9752$~\AA. The angle between the bonds is $104.52^\circ$. 
The site of the oxygen is neutral while a fourth site carries the
negative charge of the oxygen $q_O=-2q_H$. This site is located 
in the same plane of the molecule at distance $0.15$~\AA\ from the 
oxygen with an angle $52.26^\circ$ from the OH bond. The intermolecular
interactions are represented by
Coulombian terms between the charged sites and a Lennard-Jones (LJ)
potential between the neutral oxygen sites.
The LJ parameters are given by
 $\sigma_{O}=3.16$~\AA\ and $\epsilon_{O}=0.65$~$ kJ/mol$.

Water is embedded in a  rigid disordered array of $N_M=6$ soft spheres.
The soft spheres interact with the water oxygen sites by means of
the potential
\begin{equation}
v_{OM}(r)=4 \epsilon_{OM} \left( \frac{\sigma_{OM}}{r} \right)^{12}
\end{equation} 
where 
\begin{equation}
\epsilon_{OM}=\sqrt{\epsilon_O \epsilon_M} 
\end{equation}
and
\begin{equation}
\sigma_{OM}=\frac{1}{2} \left(\sigma_O + \sigma_M \right) 
\end{equation}
We assume $\sigma_M=2\sigma_O$ and $\epsilon_M=0.1 \epsilon_O$.
The average distance between the centers of two soft spheres 
is $10.5$~\AA.
With this size the soft spheres are far from the limit 
where they appear as hard wall to water. On the other end 
to enhance the volume excluded effects we use a pure repulsive
water-spheres 
interaction where the parameter $\epsilon_M$ of the soft 
potential is taken as $0.1\epsilon_O$ to soften the repulsive ramp.
The order of magnitude of the size of the spheres is close to the 
typical hard sphere diameter of apolar solutes~\cite{chandler2,graziano}. 

The simulations are performed in the $NVT$ ensemble with the use of 
a Berendsen thermostat. The potentials are truncated at $9$~\AA. 
The corrections to the long range Coulombic interactions are taken
into account with the Ewald method. The timestep used is $1$~fs.

The density of water $\rho$ is  calculated on assuming
that the excluded volume due to the confining spheres is approximately 
$V_{excl}=\pi N_M \sigma_{OM}^3/6$, in this way for a given number of water
molecules $N_w$ 
\begin{equation}
\rho=\frac{N_w}{L^3-V_{excl}} \frac{N_A}{W_{mol}}
\label{rhow}
\end{equation}
where $L$ is the boxlenght, $N_A$ is the Avogadro number and 
$W_{mol}$ is the molar volume. 
In the starting configuration the water molecules 
are arranged with the center of mass in a cubic lattice configuration where
there are voids corresponding to the random positions of the matrix spheres.
In order to have enough space for the spheres the boxlength is fixed to
$L=19.25$~\AA. Then for a given density
we place in the box a number of water molecules according to Eq.~\ref{rhow}.
The range of temperatures spanned is $200$~K $<T<350$~K and the 
density range is $0.75$~$g/cm^3 <\rho< 1.02 $~$g/cm^3$.
For each density the system is melted at $T=500$~K and equilibrated 
at several temperatures in the range indicated above. 
The longest runs, corresponding to the
lowest temperatures investigated, lasted $10$~ns.
The soft spheres positions are the same for all the states simulated.
We checked for an isochore that changing the disordered configuration
of the soft spheres did not significatively alter the pressure values
(see fig. \ref{fig:11}).

\begin{figure}[h]
\centerline{\psfig{file=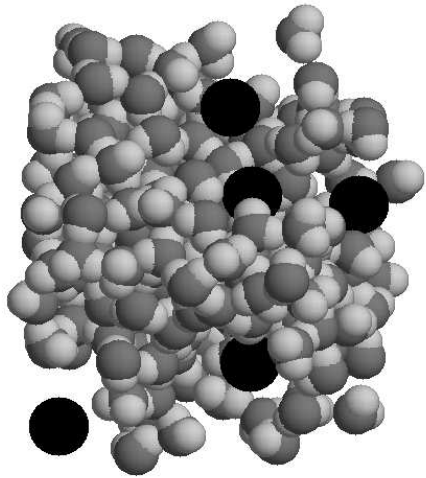,width=8.0cm}}
\caption{Snapshot of an equilibrated configuration of the system
at $T=300$~K and $\rho=0.85$~$g/cm^3$. Black spheres:
hydrophobic solute; gray spheres: oxygen atoms; lightgray spheres: 
hydrogen atoms.}
\label{fig:1}
\end{figure}

In Fig.~\ref{fig:1} we show a snapshot of an equilibrated configuration
for $T=300$~K and $\rho=0.85$~$ g/cm^3$.    
It is evident the depletion regions around each hydrophobic sphere
and the uniform distribution of water molecules elsewhere.

\section{Structural properties}

\begin{figure}[h]
\centerline{\psfig{file=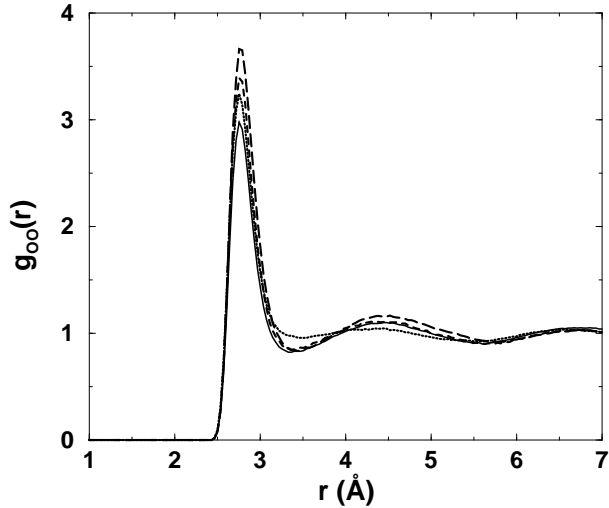,width=8.0cm,clip=!}}
\caption{Site-site pair correlation functions $g_{OO}(r)$ at
$T=300$~K for the confined system at densities $\rho=1.0$~(dotted line),
$0.90$~(dashed line), $0.75$~(long dashed line)~$g/cm^3$ compared
with the bulk at $T=300$~K and $\rho=1.0$~$g/cm^3$~(solid line).
}
\label{fig:2}
\end{figure}

\begin{figure}[h]
\centerline{\psfig{file=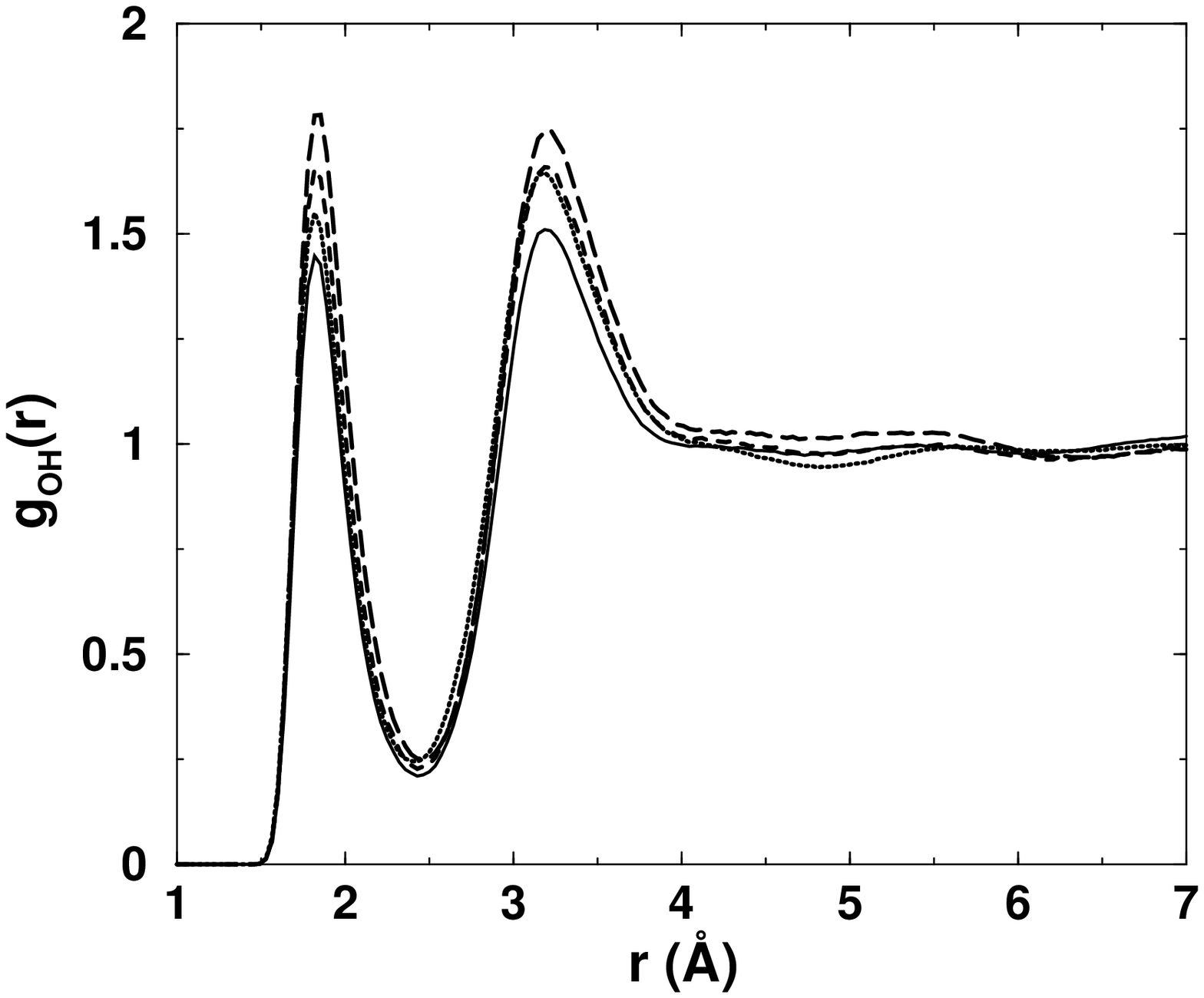,width=8.0cm,clip=!}}
\caption{Site-site pair correlation functions $g_{OH}(r)$ at the
same temperature and densities as Fig.~\ref{fig:2}.  
}
\label{fig:3}
\end{figure}

\begin{figure}[h]
\centerline{\psfig{file=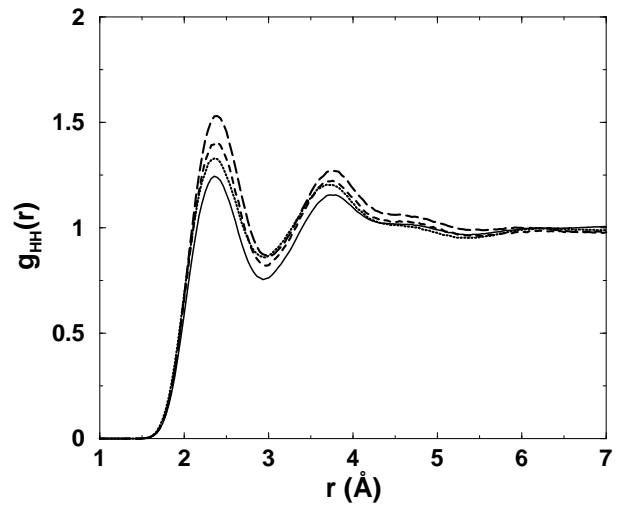,width=8.0cm,clip=!}}
\caption{Site-site pair correlation functions $g_{HH}(r)$ at the
same temperature and densities as Fig.~\ref{fig:2}
}
\label{fig:4}
\end{figure}

The site-site pair correlation functions have been calculated averaging
the equilibrated configurations to see the effect of changing the
density of water.
The results for the water-water functions
at ambient temperature are reported in Fig.~\ref{fig:2}-\ref{fig:4}
for different densities and compared with the 
bulk for $\rho=1$~$g/cm^3$. The trend is similar
to the one found in aqueous solutions of rare 
gases~\cite{guillot,vdegrandis,pcris}, 
considering
that the decrease of the density in our case is equivalent to an increase of
the solute concentration. The raise of $OO$ peak in Fig.~\ref{fig:2} 
with decreasing density is the signature of the enhancement of
the structuring effect when less molecules are present and they
can more easily approach one another. This is also evident in the
behaviour of the $g_{OH}(r)$ in Fig.~\ref{fig:3} and  $g_{HH}(r)$
in Fig.~\ref{fig:4}. 

\begin{figure}[h]
\centerline{\psfig{file=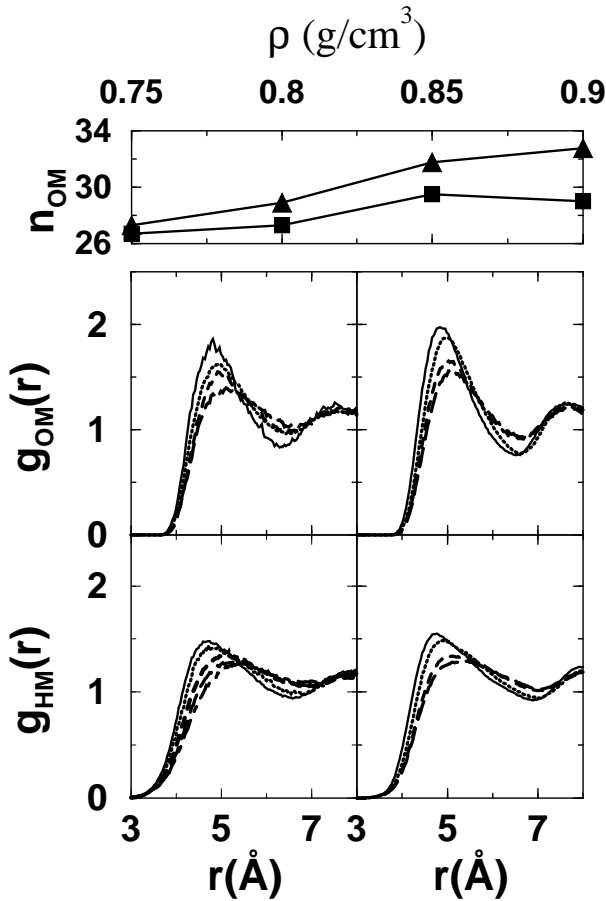,width=8.0cm,clip=!}}
\caption{Pair correlation functions $g_{OM}(r)$ (middle panel), $g_{HM}(r)$
(bottom panel) for $T=300$~K on the left and $T=220$~K on the right
and densities $\rho=0.90$~(solid line), $0.85$~(dotted line), 
$0.80$~(dashed line), $0.75$~(long dashed line)~$g/cm^3$.
In the top panel 
number of oxygen nearest neighbors $n_{OM}$ as function of density
for  $T=300$~K (full squares) and $T=220$~K (filled triangles).
}
\label{fig:5}
\end{figure}

On the contrary on decreasing density the cages formed by water 
around the confining spheres become less structured
as can be seen in Fig.~\ref{fig:5} where
the water-sphere radial distribution functions are reported
for two different temperatures and different densities.
The first peak of the $g_{OM}(r)$ becomes less pronounced and 
shifts toward higher distances, a signature of an increase of the 
hydrophobic repulsion at lower density.
In the upper inset of Fig.~\ref{fig:5}  we show 
the number of nearest neighbors $n_{OM}$ defined as follows
\begin{equation}
n_{OM}=4 \pi \rho c_O \int_0^{r_{min}} g_{OM}(r)r^2dr
\end{equation}                             
where $r_{min}$ is the value of the interatomic distance at which
the first minimum in $g_{OM}(r)$ is located, $c_O$ is the concentration
of oxygen atoms and $\rho$ is the total density.
We see that number of nearest neighbors $n_{OM}$ 
reported in the top portion of the figure goes down as 
the water density decreases. In Fig.~\ref{fig:5} it is also shown 
the effect of the decreasing temperature.  
The cage becomes more ordered and more water molecules are
present on the average in the first shell around the soft spheres 
at high density. 
On lowering the density the
$n_{OM}$ reaches an asymptotic value which appears to be independent
from the temperature.

The behaviour of $g_{HM}(r)$ also reported in  Fig.~\ref{fig:5}
shows a similar trend as function of temperature and density.
The first peak of $g_{HM}(r)$ is at the same position of the $g_{OM}(r)$
one, this indicates that the soft sphere is located interstitially
in the hydrogen bond network equidistant on the average
from the oxygens and hydrogens.
We notice in $g_{HM}(r)$ at lower temperature for the 
higher density the appearance of a shoulder after the first peak, 
indicating that some of the O-H bonds point radially toward the
second shell. 
Looking at the second peak of $g_{OM}(r)$ and $g_{HM}(r)$
we notice that the water molecules in the second shell are preferentially
oriented with the oxygen atoms towards the sphere and the
hydrogen atoms towards the water similar to the trend found in
aqueous solution of water and rare gases~\cite{guillot,vdegrandis,pcris}.

\begin{figure}[h]
\centerline{\psfig{file=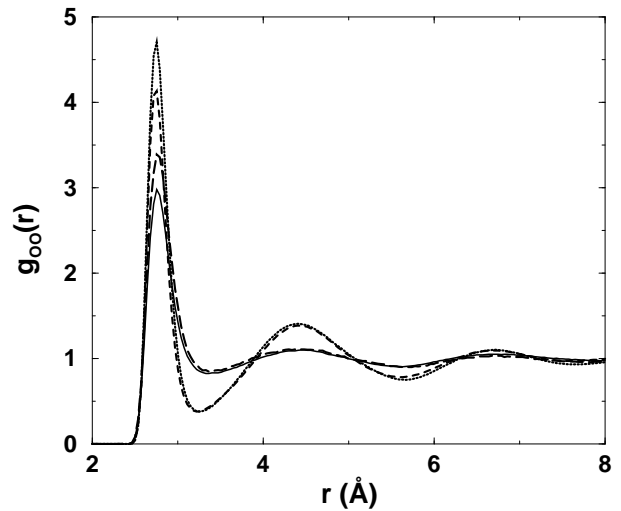,width=8.0cm,clip=!}}
\caption{
Site-site pair correlation functions $g_{OO}(r)$  for the confined system 
at density $\rho=0.90$~$g/cm^3$ and temperatures $T=300$~K (long dashed
line) and $T=220$~K (dotted line) compared with the bulk 
at density $\rho=1.0$~$g/cm^3$ and temperatures  $T=300$~K (solid line) 
and $T=220$~K (dashed line). 
}
\label{fig:7}
\end{figure}

In Fig.~\ref{fig:7} to show the effect of the 
temperature on the structure of water we report
the $g_{OO}(r)$ for $T=300$~K and $T=220$~K 
and $\rho=0.90$~$g/cm^3$. The radial function shows 
maxima and minima at the same position but more pronounced
at lower temperature. 
We notice that the $g_{OO}(r)$ for
$\rho=0.90$~$g/cm^3$ in the confined system is very similar to the 
$g_{OO}(r)$ for $\rho=1.0$~$g/cm^3$ in the bulk apart from a difference in the
height of the first peak. This similarity is preserved at decreasing 
temperatures.

To complete the analysis of the 
change of density on structure of water we consider the
changes in the H-bond network. We define the hydrogen bond by
geometric conditions. Two molecules are hydrogen bonded if the
intermolecular $O \cdots O$  distance is less than $3.35$~\AA\ and
the angle between the intramolecular $O-H$ vector and the
intermolecular $O \cdots O$ vector is 
less than $30^o$.
 
\begin{figure}[h]
\centerline{\psfig{file=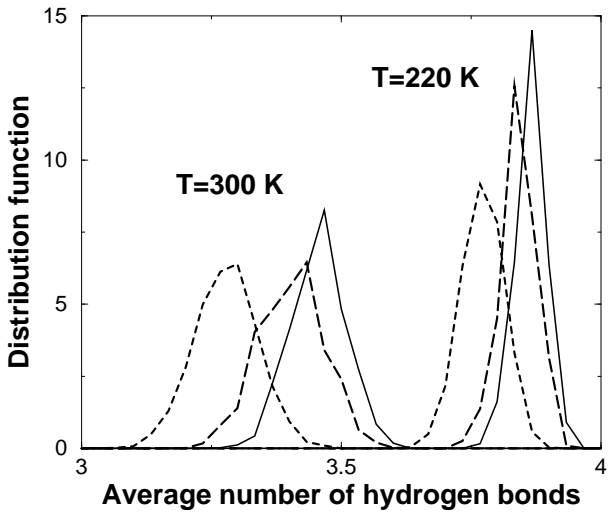,width=8.0cm,clip=!}}
\caption{ Distribution functions of the
average number of hydrogen bond per molecule at temperatures 
$T=300$~K and $T=220$~K for the bulk at density $\rho=1.0$~$g/cm^3$ 
(solid line) and for the confined system
at densities $\rho=0.90$~$g/cm^3$ (long dashed line) and 
$\rho=0.75$~$g/cm^3$ (dashed line). }
\label{fig:8}
\end{figure}

In Fig.~\ref{fig:8} it is reported the distribution function of the
average number of H-bond per molecule at two different temperatures and
three densities. We have computed the average HB number for each 
istantaneous configuration and then we have calculated the
distribution function of these average values over many configurations. 
We now discuss the results for the bulk at 
normal density compared with $\rho=0.90$~$g/cm^3$ 
in the confined system since they
appear to have a similar structure as shown above.
It is evident that decreasing water density in
confinement has the
effect of reducing the average number of hydrogen bonds in
comparison with the bulk even when $g_{OO}(r)$
are similar. At ambient temperature in confinement
the distribution becomes broader and its peak position shifts to lower
value with respect to bulk. These effects are enhanced by
decreasing the density. At lower temperatures the H-bond distributions
becomes sharper while the peak position moves toward a value close to $4$  

\begin{figure}[h]
\centerline{\psfig{file=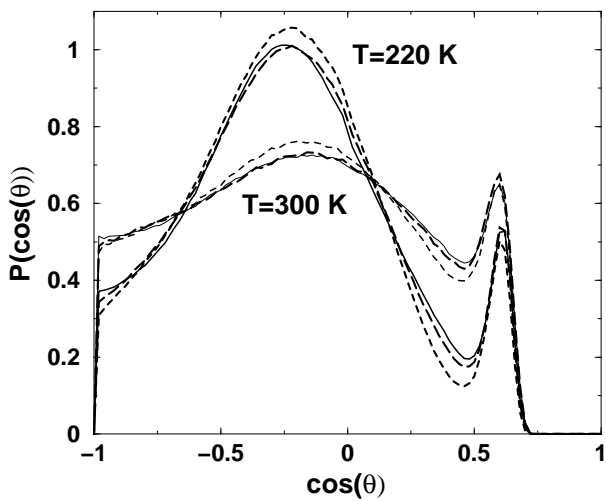,width=8.0cm,clip=!}}
\caption{Distribution functions of 
$cos(\theta)$ (see text for definition) at temperatures 
$T=300$~K and $T=220$~K for the bulk at density $\rho=1.0$~$g/cm^3$ 
(solid line) and for the confined system
at densities $\rho=0.90$~$g/cm^3$ (long dashed line) and 
$\rho=0.75$~$g/cm^3$ (dashed line). 
}
\label{fig:9}
\end{figure}

The confinement however preserves the geometry of the H-bond network.
This can be deduced from the distribution of  
the angle $\theta$ formed by the vectors
joining the oxygen atom of a central molecule and
the oxygens of two nearest neighbors water 
molecules which are H-bonded.
This distribution is reported in  Fig~\ref{fig:9}. It is 
evident that the trend is very similar between the 
bulk and the confined case. On lowering the temperatures
the H-bond network approaches more closely the tetrahedral
ordering with a sharper
distribution around the angle $104^o$ of the TIP4P
potential, while
the number of interstitial molecules decreases.
We note that the confined $\rho=0.90$~$g/cm^3$ system has a distribution
similar to the bulk $\rho=1.0$~$g/cm^3$ system. 

In analogy with previous work on water confined
between hydrophobic plane walls~\cite{starr,torquato} we find a reduction 
of the average number of H-bond. 
In our case
the H-bond network does not appear to be deformed by the presence
of the matrix. In the hydrophobic shells around the soft spheres
water molecules maintain the H-bond
network, but since they cannot form H-bonds with the neutral
soft spheres the average number of hydrogen bonds is smaller
compared with the bulk. 
As a general trend we observe that confined water  
seems to be equivalent to bulk water 
at an higher density (see Fig.~\ref{fig:7}), it is also evident
that there are not dramatic changes in the network apart from
a decrease of the hydrogen bond average number.

\section{Calculation of the spinodal line}

As we have shown in the previous section the soft sphere matrix does not
induce large changes in the structure of water.
We investigate now how 
the thermodynamical behaviour
of confined water is changed with respect to the bulk and
we will see that for thermodynamics important changes take place. 

To study how the confinement in the hydrophobic matrix affects
the thermodynamic stability of liquid water we calculate the
liquid branch of the isotherms of our system upon cooling. 
From the isotherms the limit of mechanical stability can be determined
in the framework of mean field theories from the divergence of the
isothermal compressibility.
The singularity points where 
\begin{equation}
\left( \frac{\partial P}{\partial \rho} \right)_T=0
\label{eq:spinod}
\end{equation}
define the spinodal line. They
can be obtained searching for the minima of the $P_T(\rho)$ curve.
We consider here the liquid spinodal. In bulk water the 
liquid spinodal~\cite{spinodal1,Essmann,poole,tanaka,hpss,netz,Mossa,minozzi}
has been found to extend to the region of negative pressures. 
\begin{figure}[h]
\centerline{\psfig{file=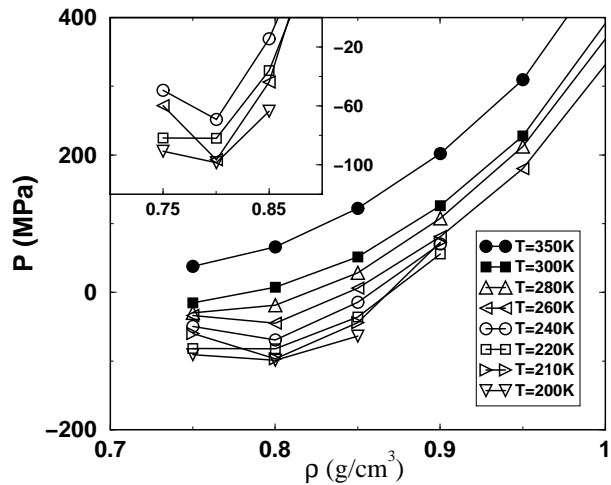,width=8.0cm,clip=!}}
\caption{Isotherms for densities ranging from $\rho=0.75$~$g/cm^3$
to  $\rho=1.0$~$g/cm^3$ for temperatures ranging from $T=350$~K to $T=200$~K.
}
\label{fig:10}
\end{figure}

\begin{figure}[h]
\centerline{\psfig{file=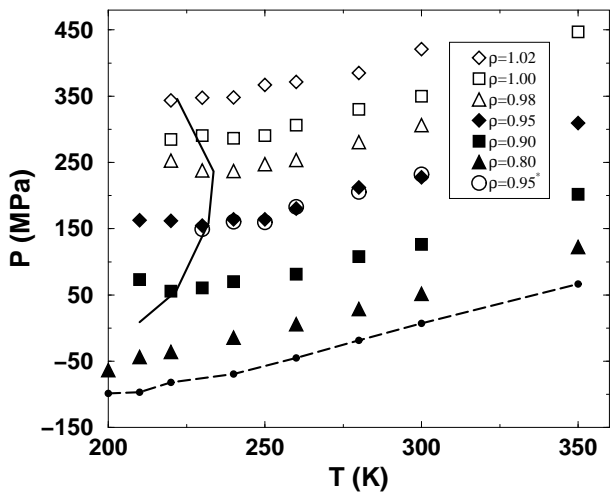,width=8.0cm,clip=!}}
\caption{$P_\rho(T)$ of the
confined system isochores for several values of $\rho$
in $g/cm^3$, spinodal line (long dashed with points)  and TMD 
curve (solid line). * Open circles represent the $\rho=0.95$ 
$g/cm^3$ isochore obtained 
from a different configuration of the matrix of 
soft spheres. }
\label{fig:11}
\end{figure}

In Fig.~\ref{fig:10} we show the isotherms of the system obtained upon 
cooling for densities ranging from  $\rho=0.75$~$g/cm^3$
to  $\rho=1.0$~$g/cm^3$ and temperatures from $T=350$~K to
$T=200$~K.
Below $T=280$~K minima start appearing in the curves in 
correspondence with $\rho=0.8$~$g/cm^3$. In the bulk~\cite{spinodal1} 
the minima appear
between $\rho=0.85-0.87$~$g/cm^3$.
From Eq.~\ref{eq:spinod} we see that the minima of the isotherms represent
the limit of stability of the liquid. The corresponding $P_S(T)$
spinodal line is plotted in Fig.\ref{fig:11} together with the 
isochores of the system. 
In  Fig.\ref{fig:11} we report for $\rho=0.95$~$g/cm^3$
isochore also 
the results obtained for a different configuration of soft
spheres. As it can be seen changing the soft spheres position
does not alter significatively the values of the pressure.
From the isochores also the behaviour of the $TMD$
line can be extracted. In fact along the TMD line the coefficient
of thermal expansion
\begin{equation}
\alpha_P=\frac{1}{V} \left( \frac{\partial V}{\partial T} \right)_P
\end{equation} 
goes to $0$.

The thermal pressure coefficient 
\begin{equation}
\gamma_V= \left( \frac{\partial P}{\partial T} \right)_V
\end{equation} 
is connected to the coefficient of thermal expansion by the following
equation
\begin{equation}
\gamma_V=\frac{\alpha_P}{K_T}.
\end{equation}

\begin{figure}[t]
\centerline{\psfig{file=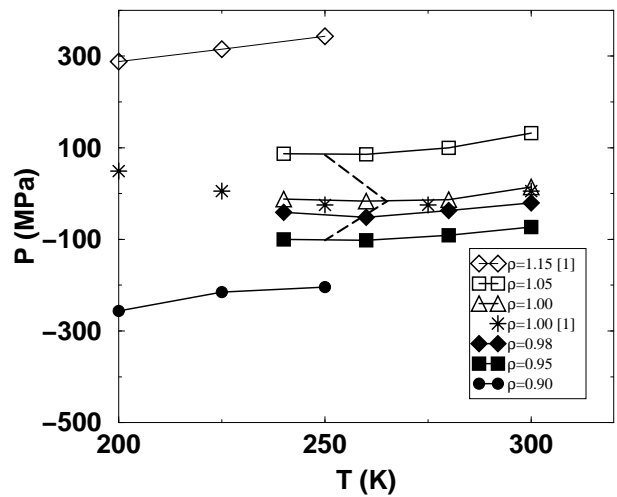,width=8.0cm,clip=!}}
\caption{$P_\rho(T)$ of the
bulk system isochores for several values of $\rho$
in $g/cm^3$, TMD curve (long dashed line). The isochore of the
lowest density is close to the spinodal. 
}
\label{fig:12}
\end{figure}

\begin{figure}[t]
\centerline{\psfig{file=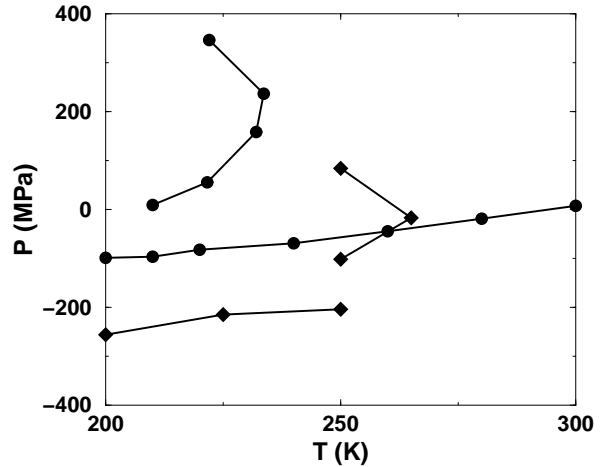,width=8.0cm,clip=!}}
\caption{TMD and spinodal of the confined system (bold lines with
filled circles) and TMD and isochore of the
lowest density for the bulk system (bold lines with filled diamonds).}
\label{fig:13}
\end{figure}

Therefore the TMD points lie on the line connecting 
the minima of the isochores.
In the Fig.~\ref{fig:11} also the TMD line is drawn.
We observe, down to the lowest temperature investigated a non retracing
spinodal and correspondingly a TMD that bends on approaching the spinodal.
This behaviour is similar to both non polarizable~\cite{spinodal1} and 
polarizable~\cite{minozzi} bulk TIP4P water. 
The interesting feature is represented
by the shift induced by confinement on both the spinodal and the TMD
with respect to the bulk. To give a more quantitative insight on this
displacement we plotted in Fig.\ref{fig:12} the isochores and the TMD
for the bulk TIP4P. Some of the isochores have been calculated here
as they are not available in literature. By comparing the 
spinodal and TMD curves of the bulk and confined system,
Fig.\ref{fig:13},  
we note  that both the spinodal line and the TMD for
water under hydrophobic confinement are shifted toward
higher pressures. The TMD line shifts also to lower temperatures
and correspondingly its curvature broadens. In a different 
hydrophobic confinement~\cite{starr}, as already quoted in the introduction,
a negative shift of $40$~K has been observed with respect to the bulk
for the TMD, but
no shift in pressure nor changes in the curvature were found.
Interestingly our shift in temperature is also roughly $40$~K.  

\section{Conclusions}

We performed a MD study of TIP4P water confined in a rigid matrix
of soft spheres. 
The geometry of the hydrophobic environment in our case
is different with respect to the previous studies mentioned above where water 
is confined between plane surfaces.
We observed that the effect on water structure 
is similar to the case of a solution of small
apolar solutes when it is taken into account that the decrease
of water density is equivalent to an increase of solute 
concentration. Moreover we  found that
volume excluded effects reduce the 
average number of hydrogen bond,
The hydrogen bond network however is preserved at variance
with the case of water confined in hydrophobic plates.

In spite of the preservation of the network of water and
the not large changes in its structure 
we observe an important shift of the limit of stability both 
in pressures and temperatures and also a different shape in the
shifted TMD with respect to the bulk.  
The temperature shift similar to the one
observed by Kumar et al.~\cite{starr}. These simulation
indicate only a weak dependence of water properties on
the confining potential~\cite{Starr2}.
It would be therefore
valuable to understand to what extent an analogy can be drawn.

\end{document}